# Measuring the Success of Software Process Improvement: The Dimensions

Pekka Abrahamsson
*University of Oulu, Finland*
Pekka.Abrahamsson@oulu.fi

## Abstract

Quality managers, change agents and researchers are often troubled in defining and demonstrating the level of success achieved in software process improvement (SPI) initiatives. So far, there exist only few frameworks for identifying the level of success achieved in SPI. Analysis shows that these frameworks do not provide a comprehensive view from all relevant stakeholders involved in SPI. Early results from an ongoing research effort to discover and operationalise success dimensions are reported. Adapted from the project management literature it is suggested that five dimensions characterise the level of success achieved in SPI: (1) project efficiency, (2) impact on the process user, (3) business success, (4) direct operational success and (5) process improvement fit. Results from an empirical analysis are reported where 23 change agents evaluated the relative level of importance of each dimension. Early results indicate that change agents valued the process user satisfaction the most and the process improvement fit the least. This finding confirms the need of having various stakeholders and dimensions acknowledged in a framework that is used to measure the overall success of an SPI initiative.

## Introduction

In recent years many excellent books on practical software process improvement (SPI) (e.g. [8, 10, 21]) have been written. Still change agents, quality managers, process owners and researchers are often troubled in defining the level of success achieved in SPI.

There does not exist a universal framework with which the success of a project could be measured and assessed [18]. As of today, this argument remains true in engineering like projects as well as process improvement projects. However, some authors (e.g. [15]) call for learning from the previous successes while others (e.g. [1]) call for learning from failures. In order to learn from previous experiences, one needs to be able to separate a success from a failure. SPI research has paid little attention to systematic study about the conditions under which SPI initiatives vary in their results [4]. The issue of success of an SPI initiative is deemed to be a problematic one since we all have experiences on projects that initially appeared to be failures (e.g. projects were not finished in time) but later on turned out to be successes (e.g. fault rate was dramatically dropped).

It has been acknowledged that software measurement is essential in the improvement of software processes and products since if the process (or the result) is not measured the SPI effort could be addressing the wrong issue [21]. However, the issue of process measures is generally considered to be a level four issue in the widely used Capability Maturity Model (CMM[1]). Still, vast majority of the organisations undertaking SPI activities are in level one or two in respective scale. For a such type of an organisation the undertaking of process measures could prove to be a difficult task since in spite of extensive literature on software measurement (see e.g. [16] for overview) companies are facing serious problems initiating even the simplest metrics programs [9].

The issue of direct measurement of a success of any particular SPI initiative is not the main objective of this paper however, since a measure developed without thorough understanding of the concept of interest is not a true measure and may result e.g. in serious ambiguities when interpreting results [20]. Therefore, as a step toward a mature measurement of success in SPI the dimensions that can be used to characterise success are introduced. These dimensions are drawn from project management literature and adapted to SPI environment. It is shown that any direct measure of success remains inadequate if other dimensions are not considered. Also, it is demonstrated that the importance of these dimensions vary depending on the stakeholder (e.g. software developer, change agent or manager) evaluating it. In this paper, early results of an empirical test are reported where 23 change agents evaluated the level of importance of each dimension from their point of view. In short, results indicate that a) all process success dimensions were evaluated to be at least moderately important to change agents, b) the process user (i.e., target of change) satisfaction dimension was rated the highest while c) the ability of an SPI initiative to prepare the organisation for future initiatives was ranked the lowest.

This paper is organised as follows: First, a sample selection of studies that address the issue of success in SPI are introduced. Secondly, the concept of success in SPI is considered by (a) defining the stakeholders involved in SPI initiative, (b) introducing the dimensions that characterise success in SPI, and (c) addressing the issue of Characterising the success using the dimensions identified. Thirdly, results from an empirical study are presented together with conclusions and a future research agenda.

---

[1] Capability Maturity Model and CMM are service marks of Carnegie Mellon University

## Related Literature

El Emam, Goldenson, McCurley and Herbsleb [4] present a multivariate model of conditions that may be used to explain the successes and failures of SPI efforts. Their model is based on data of CMM based improvement initiatives. Their main finding is that the factors distinguishing between successes and failures are the extent the organisation is focused in its improvement effort, organisational politics and the level of management commitment. El Emam et al. address an important issue but by identifying factors that contribute most to the success of an SPI initiative they do not differentiate the type of successes they deal with. Fitzgerald and O'Kane [6] did an action research study and focused on identifying factors that contribute to success in achieving particular CMM maturity level. Their data is drawn from Motorola's Cellular Infrastructure Group in Cork, Ireland. While their findings are of great interest to organisations pursuing process improvement activities using the CMM approach, the concept of success remains vague to the reader. These two studies are examples of a stream of research aiming at identifying critical success factors of SPI initiatives. Others (e.g. [11, 21]) have also addressed the problem.

Recently Reo, Quintano and Buglione [17] addressed the issue of measuring SPI by extending the infrastructure and innovation perspective of ESI's (European Software Institute) Balanced IT Scorecard (BITS) Generic Model. They propose using the model in identifying goals, drivers and indicators that can be further used in aligning improvement programmes with business objectives. Their extended model addresses e.g. people issues (satisfaction and competence) and organisational climate (continuous improvement, organisational learning and innovation) issues aside the traditional technological issues. They emphasize the need for addressing a wider range of issues than directly process related when measuring software process improvement. While providing a comprehensive list of themes that affect e.g. employee satisfaction and competence, they do not address the issue of success directly other than pointing out that in order to be successful one needs to concentrate on these themes.

Kitchenham [12] has written a comprehensive book of measuring SPI. The book aims at explaining how software measurements can be used to support SPI activities. However, while addressing the measurement issues Kitchenham does not take into account other stakeholder positions. She seems to discuss software metrics as sole indicators of success in SPI. For example, process user satisfaction with the altered processes is not discussed. While hard metrics[2] do provide a great deal of support for process improvement they do not give whole-hearted picture of the impact that a particular SPI programme has on the organisation.

## Considering Success in SPI

Collins Electronic English Dictionary defines the term 'success' as "the favourable outcome of something attempted". This literary definition gives a lot of room for interpretations and invites confusion if directly operationalised without considering e.g. all viewpoints present in SPI. Therefore, we shall start considering

---

[2] Hard metrics (e.g. productivity in terms of lines of code per staff month or design size in terms of number of modules in a product) is used here as opposite to soft metrics (e.g. work morale or level of excitement).

success by introducing all relevant stakeholders that may be involved in SPI programme.

**Stakeholders in SPI**

There are multiple types of roles that different interest groups play in an SPI project such as the sponsorship role, management role, coordination role, improvement team role (operational role) [21] and the end-user role (the target of change). Within each of these roles there are people involved with different professional and personal aims as well as different ambitions.

Success means different things to different people [7]. The corporate executive acting in the sponsorship role is concerned with the overall organisational business benefits yielding from the process improvement initiative while on the other end of the spectrum the targets of change (e.g. software developers) might be more concerned with the usefulness of the new or modified process expecting to benefit from the new tool, or procedure in a way that enables them do their jobs better. Targets of change as end-users of SPI activities are in a key position when considering the success of SPI. In general, a systems development project is considered to be a failure if the system developed is not used [14]. Similarly in SPI, if the improved process is not used after initial trials, it should be considered as a failure. Any type of software process improvement success measurement should therefore reflect a more of a multidimensional approach reflecting the expectations of senior management as well as targets of change rather than a single dimensional approach. Table PABRAHAM.1 summarizes the role[3] of various stakeholders and their function in an organisational software process improvement programme.

| Role | Person(nel) | Function in SPI |
| --- | --- | --- |
| Sponsorship | A corporate executive | Authorization of budgets and resources |
| Management | A steering committee | Provides management guidance and strategies; monitors progress; resolves organisational issues; promotes SPI goals |
| Coordination | SEPG (Software Engineering Process Group) | Provides coordination; technical guidance; owner of the SPI plan |
| Operational | Software process improvement team (change agents) | Manages and implements process improvement activities |
| Object of change | E.g. software developers, testers | Participates in the change; adopts new behaviours or tools; alters their attitudes |

Table PABRAHAM.1: Stakeholders typically involved in software process improvement programmes [21]

---

[3] Depending on the size of an organization these roles may overlap and be performed by the same person.

## Success Dimensions

The following introduction of dimensions that can be used to characterise success in SPI is adapted from project management literature where the issue of success has been under debate for over two decades. Project management literature can be seen as a suitable reference literature since most SPI activities are performed as projects with their own budget, timescale, dedicated resources and goals. The empirical part of the study identifies the importance of these dimensions from the viewpoint of change agents. Change agent is viewed here as a person or a group of persons (operational role in SPI, in Table PABRAHAM.1) facilitating and/or responsible for the attempted change effort.

Shenhar, Dvir and Levy [18] have proposed a multidimensional, multicriteria success assessment framework, which in their view reflects different interests and views set by various parties involved in the project. Shenhar et al. provided a set of dimensions that accumulate the total success level of a project. Each of the following dimensions is important in their own right depending on the type of improvement initiative and the extent of change pursued:

### D1: Project Efficiency (Meeting Constraints)

According to Shenhar et al. the first dimension reflects a short-term measure that expresses the efficiency with which the project was managed. This classical dimension found in all of the software project management literature in mainly concerned whether the project was (a) finished on time within (b) specified budget limits. This dimension is possible to be measured during the project (Fig. PABRAHAM.1) and the results are fully applicable for further analysis after the project has been completed. Zahran [21] lists a comprehensive list of project related measures such as the development time, slippage, work completed, effort expended, funds expended, etc. These measures can be used as post mortem (as well as ongoing) project analysis to provide feedback on the level of success in terms of project efficiency success dimension. Even though it is important to coordinate and manage an SPI initiative according to good project management principles, as of its own it does not indicate whether process improvement has been successful or not.

### D2: Impact on the Process user (object of change)

Shenhar et al. analyse the success in general level within all types of engineering projects and they refer in this case to the customer. In the case of improving software processes the customer is the user of the process (e.g. project; software developer, tester). The level of success is characterised therefore in terms of level of satisfaction with the new process, fulfilment of the needs of the software developers, solving the problem they experienced and whether the improved process actually is used.

The success in this dimension (Fig. PABRAHAM.1) may be seen relatively early in an SPI initiative as process users become more educated about the process improvement initiative. Improved work morale, excitement and positive attitudes are signs of success in the early phases of the project. Process improvement activities (e.g. coaching, information sharing and training) may reinforce process users' professional competence, which in turn are visible signs of success later on in SPI initiative. Measuring the extent this dimension is a success is more problematic than the first dimension where hard numbers are a relatively straightforward issue.

### D3: Business Success and D4: Direct Operational Success

The third dimension in Shenhar's et al. classification addresses the immediate and direct impact the project has on the organisation. In the case of improving software processes this dimension is divided into two sub dimensions concerned with (a) directly process-related aspects and (b) business-related aspects. Zahran [21] includes in his list of process-related measurement many items that directly indicate the level of impact the process improvement project has on the organisation. These measurement examples are e.g. number of change requests, amount of rework, number of process defects, cycle time measures and a maturity level.

Zahran [21] among others argue that one of the major obstacles to the adoption of process improvement is the reluctance of business management to invest in SPI. Zahran points out that this obstacle is raised from the fact that there is a general lack of reliable information on the business benefits of software process improvement. Therefore, by connecting SPI goals to business goals of the organisation the business management is more likely to invest in SPI. Such high-level strategic business goals are e.g. 'gain larger market share', 'reduce time-to-market', 'improve product quality' and 'increase the productivity'. Process improvement programme is an ongoing activity where the immediate results are connected more on the process-related aspects when the business-related results are visible much later on in SPI project's life cycle (Fig. PABRAHAM.1).

**D5: Process improvement fit and preparing for the future**

The last dimension identified by Shenhar et al. addresses the issue, using their terminology, of preparing the organisational and technological infrastructure for the future. In the case of software process improvement this preparation for the future implies to applicability of an SPI initiative to provide opportunities for future improvement activities. Companies that have had bad experiences in the past with process improvement projects are reluctant to initiate such activities again. The reasons for the failure are manifold and not many times analysed deeper as to what happened. These failing SPI projects did not prepare themselves or the organisation for the future even though they would have had many opportunities to do so. A good example of a post mortem analysis process in software development project that is applicable to SPI initiative as well is introduced in [3].

Indeed, one of the powerful mechanisms for an effective process improvement support infrastructure is its continuously improving characteristic. This requires the establishment of effective feedback mechanisms that ensure direct communication channels between process users, process owners, process groups, project managers and business managers [21]. This in turn enables the organisation to redefine its goals (in respect to SPI programme) and tune the activities depending on the feedback received from various stakeholders. Failure in one specific dimension may prepare the organisation for future challenges but only if it is taken carefully into consideration e.g. in the form of knowledge transfer. On the other hand, a success in one specific SPI action may not necessarily be as relevant as how well the action fits in the larger view of the whole set of activities.

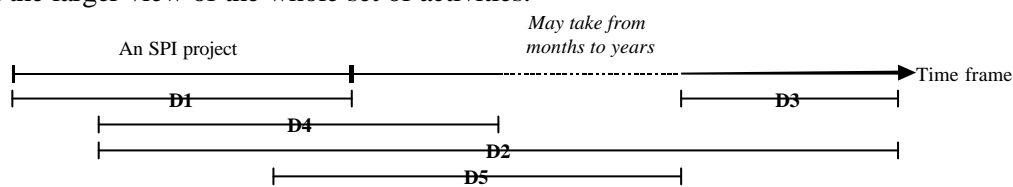

D1: Project efficiency
D2: Impact on the process user
D3: Business success
D4: Direct operational success
D5: Process improvement fit and preparation for future

Fig. PABRAHAM.1 : Success Dimensions' achievability in relation with SPI project's time frame

## The Use of Dimensions in Measuring Success

How can the success of an SPI be assessed using the dimensions identified in this paper? Is a success in SPI a simple of sum its dimensions?

Theoretically, when following Law, Wong and Mobley's [13] taxonomy of multidimensional constructs there are three possibilities for determining the relationship between the dimensions and the construct: 1) latent model, 2) aggregate model and 3) profile model. The first two shall be addressed here.

If 'success' were to belong to the latent model, its dimensions would simply be different forms (of successes) manifested by the construct. Success would therefore be a higher-level construct that underlies its dimensions. When following closely Law et al.'s argumentation on Farmer et al.'s [5] multidimensional construct on upward influence tactics, one could conclude that there exists a general multidimensional construct called 'success in SPI' which underlies five distinct dimensions that are also multidimensional constructs themselves. This line of reasoning could be confirmed when analysing the method by which Shenhar et al. [18] arrived to their success dimensions. Originally Shenhar et al. identified 13 success measures from the project management literature and from the initial phases of their study. Initially these measures were grouped into three dimensions but in their second phase of the study where total number of 127 projects were analysed factor analysis results suggested four distinct underlying dimensions (these dimensions were used as basis for defining success dimensions in SPI project). Therefore in their study, Shenhar et al. defined four success dimensions as first-order factors underlying the 13 measurement items. These four success dimensions are multidimensional constructs themselves where overall success in general is the high-order abstraction.

Latent model classification, however, suggests that any of the first-order factors (e.g. project efficiency) would equally manifest success in SPI. Theoretically speaking this could be possible but in operational use the results would be confusing and misleading since any single dimension diverges the view on the level of success if others are totally left out of the consideration. Therefore, the second model of multidimensional constructs – aggregate model suggested by Law et al. [13] is more appropriate when considering the operationalisation of the concept of success. Aggregate model enables us to determine a formula by which the level of success may be determined. Under the aggregate model classification the 'level of success achieved in SPI' is formed as an algebraic composite of its five dimensions. Weights may be used to stress the importance of certain dimensions over the others. For example, for some stakeholders the level of success in project efficiency may not be as important as achieving the business goals. Therefore, more weight should be placed on the business success dimension than on the project efficiency dimension in the evaluation process. This may prove to be difficult however, since some dimensions are more difficult to measure than others. As shown earlier in the Fig. PABRAHAM.1 all success dimensions cannot be assessed simultaneously. Therefore we label success dimensions as issues that characterise success in SPI. Future research efforts should tackle the measurement of each dimension. Tentatively it is hypothesized that the dimensions have the following (see Table 2) characteristics in terms of measurements.

|    | **Success Dimension** | **Type of measurements** | **Relative estimation difficulty** |
|----|----|----|----|
| D1 | Project Efficiency | Hard measures (e.g. work effort) | Low |
| D2 | Impact on the Process User | Soft measures (e.g. satisfaction, ease of use; work morale; level of excitement) | High |
| D3 | Business Success | Hard measures (e.g. productivity) | Moderate |
| D4 | Direct Operational Success | Hard measures (e.g. defect ratio) | Moderate |
| D5 | Process Improvement Fit and Preparation for Future | Both (experience database) | High |

Table PABRAHAM.2 : Measuring success dimensions

# Results From the Empirical Analysis

### Importance of Success Dimensions

A total of 23 change agents from 11 different organizations volunteered to evaluate the relative level of importance of each dimension from their viewpoint. Change agents were software developers, designers and process improvement staff who have responsibilities for process improvement activities. A 7-point Likert scale ranging from very low importance to very high importance with 10 items was designed to rate the importance of five success dimensions. A possibility to answer 'I don't know' was given separately apart from the 7-point scale for each item. A sample item of the first dimension (project efficiency) is illustrated below.

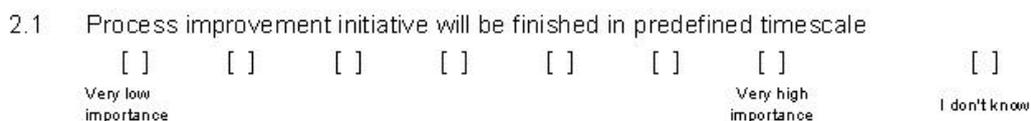

Fig. PABRAHAM.2 : A sample item from the questionnaire

Early results (Fig. PABRAHAM.3) indicate that change agents value all assessed categories to be at least moderate level of importance in regard to evaluating the success of an SPI initiative. Shenhar et al. [18] found out that project managers place greater importance to the customer (in engineering projects) than they attribute to commercial success or any other project's impact on the organisation. Similarly in the present study the change agents evaluated the process user satisfaction (dimension 2) having the highest importance above all other items or dimensions. Indeed, the importance of people issues in improving software processes has been acknowledged since only three of 23 (13%) respondents evaluated the process user satisfaction as an indicator of process improvement success lower than high importance (below six in respective scale).

Somewhat contradicting finding with the SPI literature is the result that achieving the business goals together with process improvement initiative's fit to overall SPI programme and its ability to prepare the organisation for future improvement initiatives were ranked having the lowest importance from change agent's point of view. This finding provides preliminary support for the idea that any measurements considering success in SPI should incorporate more than one single viewpoint in order to provide an adequate view on process improvement initiative's success.

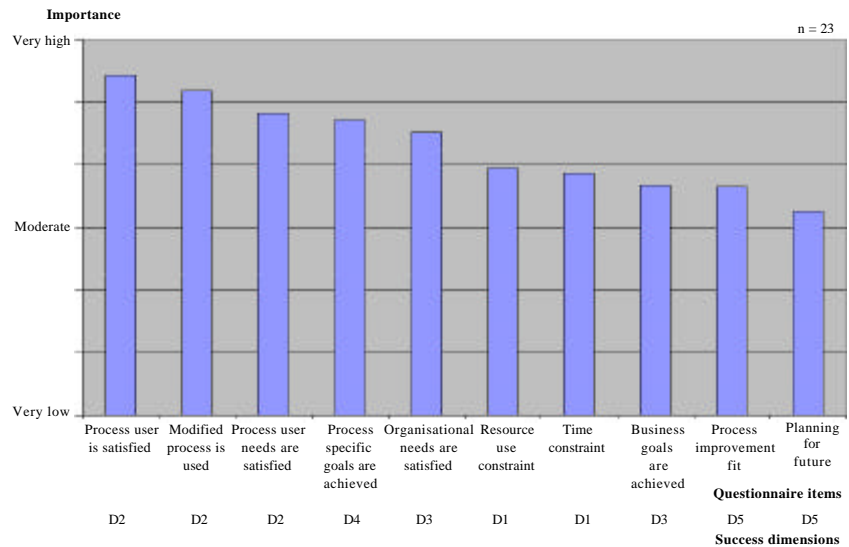

Fig. PABRAHAM.3 : The relative importance of success dimensions in SPI from the change agent's viewpoint

## Evaluating Overall Success

Even though, as indicated earlier in the paper, the issue of direct measurement is not the main concern of this paper, for practical reasons[4] the change agents (17 of them) were to evaluate the level of success they achieved from their point of view in recent SPI initiative they were responsible for. In addition changed agents evaluated the overall success they achieved using a 7-point scale.

As suspected, direct business benefits (illustrated below, in Figure 4) were the most difficult to assess since only 41 % (7 of 17) did attempt to give an evaluation of it. The 'easiest' (or the most frequently assessed) one to assess was the project efficiency dimension (D1).

---

[4] It was also studied whether the level of management commitment demonstrated would influence the success level achieved. Results are reported in [2].

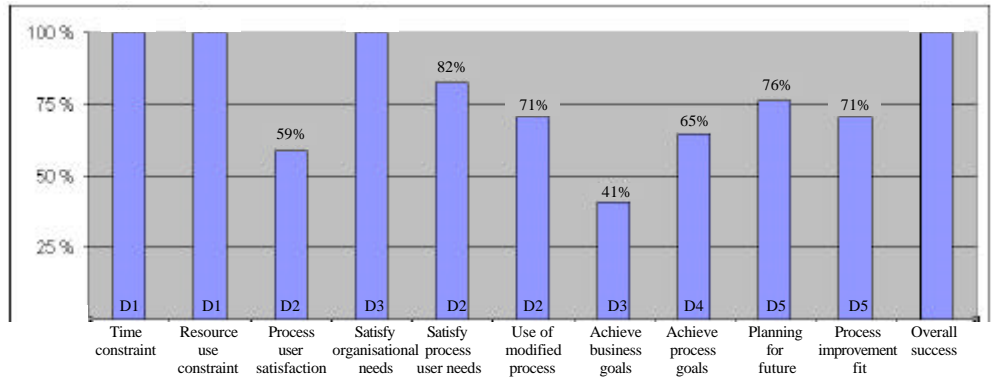

Fig. PABRAHAM.4 : Percentage of respondents per success item

To our surprise, however, there was a remarkably small difference between averaged level of achievement value (calculated as a simple sum) and weighed average (the level of importance was taken into consideration for each item) (see Fig. PABRAHAM.5). Also, the self-evaluated total level of success corresponds rather well to calculated values. Only in three occasions there is a noticeable difference. One should note that the higher the number of items assessed is, the higher should be the reliability[5] of the assessed level of achievement. It is interesting to note that the first four evaluations were done for the same SPI initiative by four different change agents. While the results (achievement levels) are extremely similar, the number of items used for evaluation varied considerably (from 4 to 10). This calls for attention on the subjectivity of the assessments when there is little data behind the evaluations.

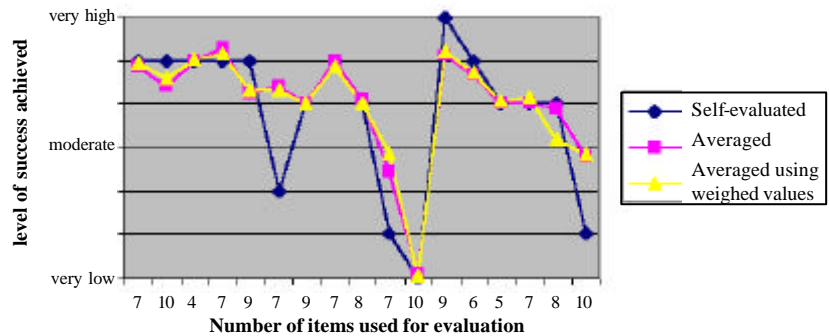

Fig. PABRAHAM.5 : The level of success achieved in SPI

## Conclusion

Practice has demonstrated that change agents, quality managers, process owners and researchers are often troubled in defining the level of success achieved in SPI. Analysis showed that existing literature on SPI success has not considered all relevant stakeholders and different levels of successes. To fill this gap five success dimensions that can be used to characterise success in software process improvement initiative were introduced. These dimensions were (1) project efficiency, (2) impact on the process user, (3) business success, (4) direct operational success and (5) process improvement fit. Change agents evaluated the relative level of importance from their point of view and the results indicated that a) all process success dimensions were evaluated to be at least moderately important to change agents, b) the process user satisfaction dimension was rated the highest while c) the ability of an SPI initiative to prepare the organisation for future initiatives was ranked the lowest.

---

[5] The quality of any assessed item depends on the quality of the data that was used as a basis for the evaluation. If all values are just best guesses, then the overall achievement level is also 'only' a best guess.

Having the dimensions identified, practitioners may find them useful e.g. in popular GQM method where metrics are derived from agreed goals by formulating targeted questions (see details in [19]). Dimensions point to the direction of the type of questions that should be asked when considering success in an SPI. Future research will concentrate on operationalising and validating suggested success dimensions. This paper reported the importance of each dimension from the change agent point of view. This view will be expanded to include other stakeholder views also. Shenhar et al. [18] proposed that achievability of various success dimensions varies depending on the type of project. Therefore, effort will be expanded on categorizing various types of software process improvement activities into a set of categories reflecting differences in the extent and type of change intended. Also, the complex set of relations between the dimensions should be defined and validated. These research activities should enable us in defining more rigorously measures that explain proposed success dimensions.

We have taken a step toward a mature measurement of success in SPI. Hopefully practitioners find these dimensions useful in explaining the level of success achieved in their software process improvement programmes. Additionally, having introduced the dimensions both researchers and practitioners are able to communicate failures and successes of process improvement initiatives using the same typology. This enables bridging the gap between science and practice and helps unifying lessons learned reports from the industry.

## Acknowledgments

Author would like to express his gratitude to Dr. Shenhar at Stevens Institute of Technology for providing the material on his study, Dr. Seppänen at University of Oulu, Ms. Parviainen at VTT Electronics and Dr. Kilpi at Nokia Mobile Phones for their insightful comments on the earlier version of this paper.